\documentclass[conference]{IEEEtran}
\IEEEoverridecommandlockouts
\usepackage{cite}
\usepackage[nolist]{acronym}
\usepackage{amsmath,amssymb,amsfonts}
\usepackage{amsmath, mathtools}
\usepackage{algorithm}
\usepackage{algorithmic}
\usepackage{graphicx}
\usepackage{textcomp}
\usepackage{xcolor}
\usepackage{braket}
\usepackage{subfig}
\usepackage{stfloats}
\def\BibTeX{{\rm B\kern-.05em{\sc i\kern-.025em b}\kern-.08em
    T\kern-.1667em\lower.7ex\hbox{E}\kern-.125emX}}
\begin{document}

\title{Quantum Deep Learning for Massive MIMO User Scheduling}

\author{\IEEEauthorblockN{Xingyu Huang \IEEEauthorrefmark{1}, Ruining Fan \IEEEauthorrefmark{1}, Mouli Chakraborty \IEEEauthorrefmark{3}, Avishek Nag \IEEEauthorrefmark{4}, Anshu Mukherjee \IEEEauthorrefmark{2}\IEEEauthorrefmark{1}}
\IEEEauthorblockA{\IEEEauthorrefmark{1} Beijing-Dublin International College, Beijing University of Technology, Chaoyang, Beijing, China\\
\IEEEauthorrefmark{2} School of Electrical and Electronic Engineering,
University College Dublin, Belfield, Dublin 4, Ireland\\
\IEEEauthorrefmark{3} School of Natural Sciences, Trinity College Dublin, The University of Dublin, College Green, Dublin 2, Ireland\\
\IEEEauthorrefmark{4} School of Computer Science,
University College Dublin, Belfield, Dublin 4, Ireland\\
Email: xingyu.huang@ucdconnect.ie, ruining.fan@ucdconnect.ie, moulichakraborty@ieee.org, \\ avishek.nag@ucd.ie, anshu.mukherjee@ieee.org}}

\maketitle

\begin{abstract}
We introduce a hybrid \ac{QNN} architecture for the efficient user scheduling in 5G/\ac{B5G} massive \ac{MIMO} systems, addressing the scalability issues of traditional methods. By leveraging statistical \ac{CSI}, our model reduces computational overhead and enhances spectral efficiency. It integrates classical neural networks with a variational quantum circuit kernel, outperforming classical \acp{CNN} and maintaining robust performance in noisy channels. This demonstrates the potential of quantum-enhanced \ac{ML} for wireless scheduling.
\end{abstract}


\section{Introduction}
The rapid expansion of 5G and \ac{B5G} technologies has significantly improved mobile service quality, leading to a surge in connected devices (30 billion by 2022). This growth has pushed traditional wireless schemes (\ac{TDMA}, \ac{FDMA}) to their limits, necessitating the adoption of \ac{mMIMO} systems for enhanced spectral efficiency. However, efficient user scheduling in \ac{mMIMO} is hindered by the high overhead of instantaneous \ac{CSI} acquisition. Statistical \ac{CSI} offers a more stable and resource-efficient alternative, but it requires intelligent algorithms to process its noisy, high-dimensional data {\cite{shi2018machine}, \cite{you2015pilot}}. 

Moreover, \ac{ML} has shown promise in learning \ac{CSI} patterns, but classical models face scalability and data limitations \cite{shehzad2021dealing}. To address this challenge, we propose a hybrid \ac{QNN} model for user scheduling tasks in a specific \ac {mMIMO} system. This proposed model leverages quantum parallelism and quantum entanglement to enhance generalization and convergence \cite{LR7}. Our model features a three-layer architecture, comprising a QNN kernel wrapped by two classical neural network (NN) layers. These layers serve as classical prelayers for feature compression and a classical postlayer for decision mapping, respectively. The \ac{QNN} itself is implemented via a quantum variation circuit that can recognise and learn patterns. Furthermore, simulations under realistic Rician fading channels demonstrate its superior performance over \ac{CNN}-based benchmarks, highlighting faster convergence and reduced overhead.

The related work is summarized as follows: Firstly, in \cite{an2023deep}, a deep reinforcement learning method is applied to user scheduling under \ac{mMIMO} networks. Secondly, the authors in \cite{LR_table_2} only demonstrated that hybrid-QNNs could assist the downlink beamforming optimization. Tertiary, in \cite{LR_table_3}, the authors present a work that leverages an RL-based method for the user scheduling problem, without using \ac{QNN} for speed-up. In addition to the existing works mentioned above, our work uniquely applies the QNN to generate a user scheduling policy, assisted by an \ac{RL}-based training method. Previous works primarily focused on other aspects (e.g., problems other than dynamic user scheduling policy generation) or used purely classical approaches. To the best of our knowledge, there has been no approach to combining statistical \ac{CSI} with quantum deep learning to address the dynamic and complex nature of user scheduling in massive MIMO environments. Furthermore, the seamless integration of classical neural networks for data preprocessing and post-processing with a quantum variational circuit for feature learning is a novel architectural design in this context. Our main contributions are as follows:
\begin{itemize}
    \item We proposed a hybrid-\ac{QNN} model, trained using an RL-based algorithm, for scheduling users in a single-cell \ac{MU-mMIMO} wireless system. The \ac{RL}-based training algorithm maximizes the expected reward by adjusting the parameters of the hybrid-\ac{QNN} policy.
    \item We architect a corresponding \ac{CNN} model as the benchmark, which is also trained by the proposed RL-based method. The comparison results demonstrate that our proposed model outperforms the benchmark in terms of sum-rate under the same channel realization.
\end{itemize}
\section{System Model and Problem Formulation}
\subsection{Beam domain received signal}
Assume that a single-cell \ac{MU-mMIMO} system is operating in the downlink transmission over the correlated Rician fading channel. This system consists of one \ac{BS} with a large-scale \ac{ULA} with $M$ antennas, known as \acp{AP}, and $K$ \acp{UE} with a single antenna. Additionally, assume we only have statistical channel state information (CSI) at the base station (BS). Consider that at most $L$ ($L \leq K$) users can be served simultaneously by the \ac{BS}. According to \cite{LR_table_3}, the received signal ${r}_l$ of $l$-th UE can be represented as
\begin{equation}
    r_l = \sqrt{P_l}\mathbf{H}_l^T \mathbf{B}_l s_l + \sum_{i=1, i \neq l}^{L} \sqrt{P_i} \mathbf{H}_l^T \mathbf{B}_i s_i + w_l
\end{equation}
where $P_l = P_i = \frac{P_{total}}{L}$, $P_{total}$ being the total transmitted power at BS, $\mathbf{B}_l\in\mathbb{C}^{M\times1}$ is the beamforming vector of the $l$-th user, $\mathbf{H}_l \in\mathbb{C}^{M\times1}$ is the channel vector, $s_l \sim \mathcal{C}\mathcal{N}(0, 1)$ is the transmitted symbol at the BS, and $w_l\sim \mathcal{C}\mathcal{N}(0, \sigma_l^2)$ represents the complex \ac{AWGN} at the receiver of the user $l$.
In the correlated Rician fading channel, the channel vector $\mathbf{H}_l\in\mathbb{C}^{M\times1}$ can be defined as \cite{li2019joint}
\begin{equation}
    \mathbf{H}_l^T = \sqrt{\frac{K_l}{K_l + 1}} \mathbf{h}_{l}^T + \sqrt{\frac{1}{K_l + 1}} \mathbf{h}_{n,l}^T \mathbf{R}_l^{1/2}
\end{equation}
where $K_l$ is the Rician $K$-factor, $\mathbf{h}_{l} \in\mathbb{C}^{M\times1}$ denotes the \ac{LoS} component which is related to the \ac{AoD} from BS to the user $l$, $\mathbf{h}_{n,l}\in\mathbb{C}^{M\times1}$ represents \ac{NLoS} component with zero mean and unit variance, and $\mathbf{R}_l\in\mathbb{C}^{M\times M}$ denotes the channel correlation matrix of the \ac{NLoS} component. For the optimal beamforming vector $\mathbf{B}_l\in\mathbb{C}^{M\times1}$, it can be approximated by using the unitary \ac{DFT} matrices to obtain the eigenvectors of the channel correlation matrix, and the diagonal matrix $\mathbf{U}_l\in\mathbb{C}^{M \times M}$ which is denoted as \cite{r7}
\begin{equation}
\mathbf{U}_l = \mathbf{W}_M^H \mathbb{E} \left[\mathbf{H}_l \mathbf{H}_l^H \right] \mathbf{W}_M
\end{equation}
where $\mathbf{W}_M\in\mathbb{C}^{M \times M}$ is the unitary \ac{DFT} matrix. The ${x}_l$-th maximum diagonal element of $\mathbf{U}_l$ corresponds to the column index in $\mathbf{W}_M$ that maximizes the energy in a specific direction for a targeted \ac{UE}. The \ac{BS} selects this column as the optimal beamforming vector for user $l$. 

\subsection{Sum rate with fairness}
Maximizing the ergodic sum rate is a pivotal objective in an effective user scheduling algorithm for downlink transmission. For user $l$, the ergodic rate in bits per second per Hz based on statistical \ac{CSI} is given by Shannon's capacity and scheduling indicator vector $\boldsymbol{\xi}=[\mathbf{\xi}^1,\mathbf{\xi}^2,\dots,\mathbf{\xi}^K]$, which is denoted as \cite{LR_table_3}
\begin{equation}
    \bar{T}_l = \mathbb{E} \left[ \log_2\left(1+\frac{P_l \left| \mathbf{H}_l^T \mathbf{B}_l \right|^2 \mathbf{\xi}^l}
{\sum_{i=1,i\neq l}^{K} P_i \left| \mathbf{H}_l^T \mathbf{B}_i \right|^2 \mathbf{\xi}^i+ \sigma_l^2}\right) \right]
\end{equation}
The goal is to clarify which $L$ users among the total $K$ users are scheduled during a time interval while maximising the sum rate, which is denoted as follows \cite{LR_table_3}
\begin{equation}
    \underset{\boldsymbol{\xi}}{\text{max}} \sum_{l=1}^{K}\bar{T}_l, 
    \ \text{s.t. }  \mathbf{\xi}^l\in \{0,1\},\ \forall{l}\in \{1, \dots, K\},\ \sum_{l=1}^{K} \xi^l = L,
\end{equation}
However, $\bar{T}_l$ is intractable because the channel vector $\mathbf{H}_l$ includes \ac{LoS} and \ac{NLoS} components in complicated ways that are difficult to simulate; thus, we approximate (4) to get a simplified expression that matches the Monte-Carlo result as follows \cite{li2019joint}
\begin{equation}
\begin{aligned}
\bar{T}_l\approx\hat{T}_l =  \log_2\left(\frac{\sum_{i=1}^{K} \beta_i a_l^{({x}_i)} \mathbf{\xi}^i + 1}
{\sum_{i=1,i\neq l}^{K} \beta_i a_l^{({x}_i)} \mathbf{\xi}^i+ 1}\right)
\end{aligned}
\end{equation} 
where $\beta_i = P_i/\sigma_l^2$ denotes the transmit \ac{SNR} in linear, $a_l^{({x}_i)}$ is the ${{x}_i}$-th diagonal element of $\mathbf{U}_l$, denoting the channel gain of beamforming in the optimal transmission direction if $i = l$, otherwise, denoting the interference channel gain. Nevertheless, maximizing the ergodic sum rate neglects fairness when some users may never obtain channel resources due to poor channel conditions. To address this problem, we employ a commonly used \ac{PF} method \cite{lau2005proportional}. The problem formula now becomes
\begin{equation}
\begin{aligned}
        \quad\underset{\boldsymbol{\xi}}{\text{max}}&\sum_{l=1}^{K}\frac{\hat{T}_l(t)}{\bar{R}_l(t-1)+\epsilon}, \\
        \text{s.t. } &\mathbf{\xi}^l\in \{0,1\},  \forall{l}\in \{1, \dots, K\}, \sum_{l=1}^{K} \xi^l = L,\\
        & \bar{R}_l(t) =
        \begin{cases}\left(1 - \alpha\right) \bar{R}_l(t-1) + \alpha\hat{T}_l(t-1), 
        \text{if $\mathbf{\xi}^l$=1} \\\bar{R}_l(t-1), \text{otherwise}\end{cases}
\end{aligned}
\end{equation}
where $\hat{T}_l(t)$ is the ergodic rate of user $l$ at the time $t$, $\bar{R}_l(t-1)$ represents the historical estimated average rate for user $l$ at the time $t-1$, $\epsilon$ is a small value to avoid division by zero, and $\alpha\in (0,1)$ represents the forgetting factor of the historical rate.
\section{Proposed Hybrid-QNN Solution}
\subsection{Three-layer Hybrid-QNN Model}
In this paper, we design and implement a hybrid-\ac{QNN} model with a three-layer architecture, as illustrated in Fig. 1. The model comprises a \ac{QNN} core and is surrounded by two classical neural networks. The Classical-pre layers are responsible for preprocessing the input features, which are flattened and scaled. At the same time, the Classical-post layer serves as a mapping function, converting the output logits of the \ac{QNN} kernel into binary scheduling indicators. The kernel of this model is implemented through a quantum variation circuit, which is illustrated in Fig. 2.
\begin{figure}[!]
    \centering
\includegraphics[width=0.9\linewidth]{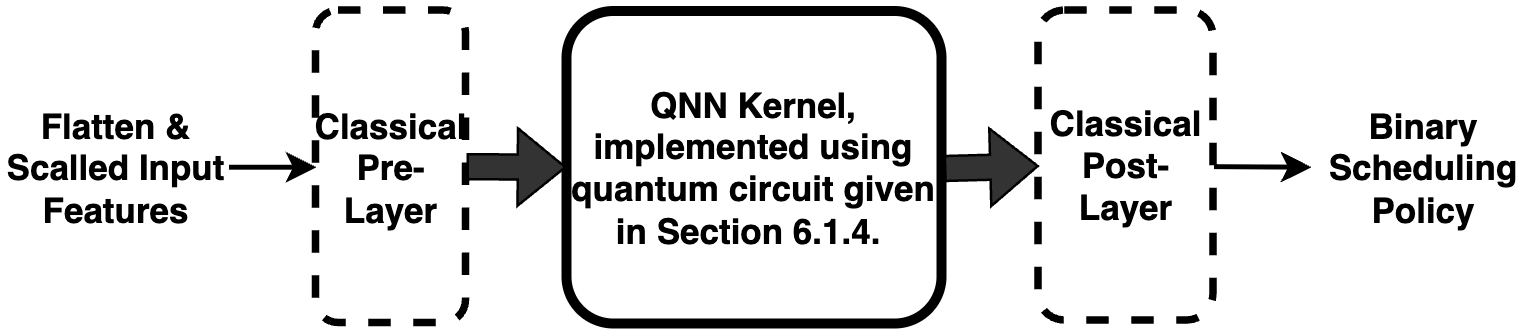}
    \caption{Three-layered Model Architecture}
    \label{fig:enter-label}
    \vspace{-0.2in}
\end{figure}

From (6), we can form an input feature matrix $\mathbf{G}_{\text{in}} \in \mathbb{R}^{K \times K}$ for each sample, denoted as \cite{LR_table_3}
\begin{equation}
    \mathbf{G}_{\text{in}} =
\begin{pmatrix}
a^{({x}_1)}_1 & a^{({x}_2)}_1 & a^{({x}_3)}_1 &  \cdots & a^{({x}_K)}_1 \\
a^{({x}_1)}_2 & a^{({x}_2)}_2 & a^{({x}_3)}_2 & \cdots & a^{({x}_K)}_2 \\
a^{({x}_1)}_3 & a^{({x}_2)}_3 & a^{({x}_3)}_3 & \cdots & a^{({x}_K)}_3 \\
\vdots & \vdots & \vdots & \ddots & \vdots \\
a^{({x}_1)}_K & a^{({x}_2)}_K & a^{({x}_3)}_K &\cdots & a^{({x}_K)}_K \\
\end{pmatrix}
\end{equation}
Then we perform data preprocessing to flatten and standardize these input features, and it is denoted as 
\begin{equation}
    \tilde{\mathbf{x}} = [\tilde{x}_1, \tilde{x}_2, \tilde{x}_3, \dots, \tilde{x}_{K^2}]^T 
\end{equation}
where $\tilde{\mathbf{x}}\in \mathbb{R}^{K^2}$ is the flattened standardized features, $\tilde{x}_{K^2}=\frac{x_{K^2} - \mu_{}}{\sigma_{}}$, $\mu$ is the mean of all input features, $\sigma$ is the standard deviation. In this paper, we apply quantum angle encoding, the most widely used encoding method due to its simplicity, where each rotation gate encodes a single real value input \cite{ovalle2023quantum}. To enable efficient angle encoding and preserve the most relevant information, we consider reducing the length of the input to the number of qubits by using a classical \ac{NN}. This process can be denoted as
\begin{equation}
    \text{input}(\tilde{\mathbf{x}}\in \mathbb{R}^{K^2})\overset{\text{classical NN}}{\longrightarrow} \text{output}(\mathbf{z}_{in}\in \mathbb{R}^{n_q})
\end{equation}
where $\mathbf{z}_{in} = [{z}_1, \ {z}_2, \ {z}_3, \dots, \ {z}_{n_q}]^T$, \( n_q \) is the number of qubits, $\mathbf{z}_{in}$ is then used as input to the \ac{QNN} kernel. The joint quantum state $\ket{\psi_{\mathbf{z}_{in}}}$ can be represented by \cite{garg2020advances}
\begin{equation}
\begin{aligned}
    \ket{\psi_{\mathbf{z}_{in}}} &=\bigotimes_{i=1}^{n_q} \ket{\psi_{i}} =\bigotimes_{i=1}^{n_q} R_Y(z_i) \ket{0}\\
\end{aligned}
\end{equation}
where $\ket{\psi_{i}}$ represents the quantum state of a qubit, $R_Y(z_i)$ denotes the rotation gate around Y-axis of Bloch Sphere. 
\begin{figure*}[!]
    \centering
    \includegraphics[width=\textwidth]{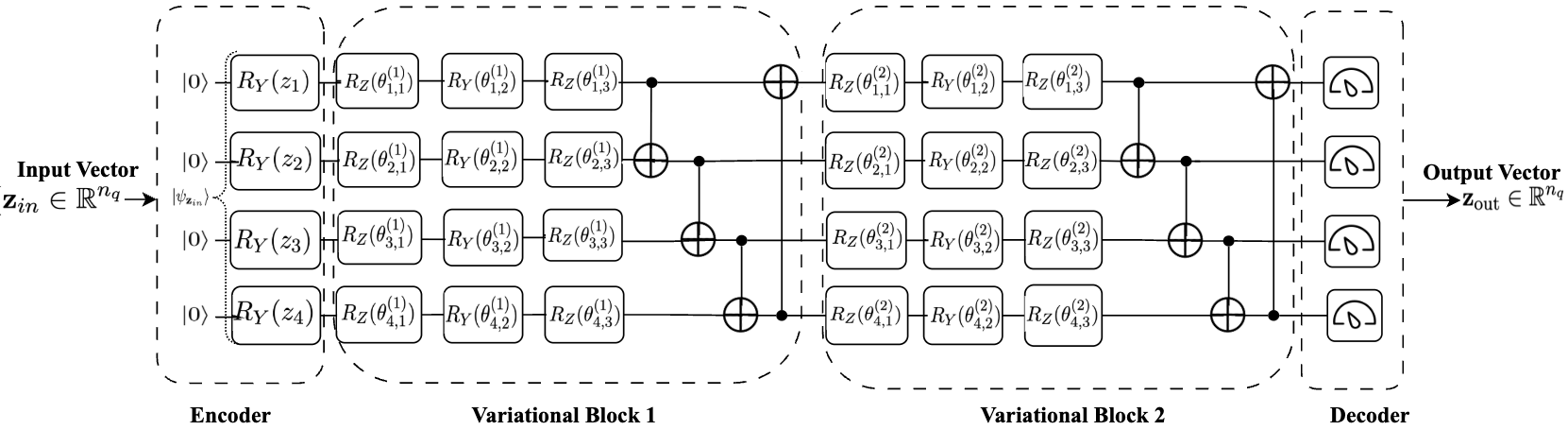}
    \caption{An example of learnable variational quantum circuits. An input vector $\mathbf{z}_{\text{in}}\in \mathbb{R}^{n_q}$ is encoded when the number of qubits is $n_q=4$ and then transformed by variational transformations. Finally, $n_q$ measurements are carried out to decode and obtain the output vector $\mathbf{z}_{\text{out}}\in \mathbb{R}^{n_q}$.}
    \label{fig:enter-label}
\end{figure*} 
In Fig. 2, the quantum gates perform unitary transformations on the encoded state $\ket{\psi_{\mathbf{z}_{in}}}$. The \ac{QNN} is most frequently modeled through learnable variational blocks, which are parametrised quantum circuits analogous to layers in classical NNs \cite{garg2020advances}. Additionally, the unparameterized \ac{CNOT} gates are used to create entanglement, allowing the \ac{QNN} to establish correlations between qubits and capture more relationships across inputs. In the quantum circuit, applying three rotation gates in the order of $R_Z$, $R_Y$, and $R_Z$ is called ZYZ decomposition, which forms a universal parametrization of arbitrary single-qubit unitaries and transmits the most information in the variational circuit \cite{krol2022efficient}. The \ac{CNOT} gates connect qubits in a ring topology, enabling \ac{QNN} to represent complex quantum correlations instead of requiring a large number of parameters or more complicated circuits \cite{deng2017quantum}. In Fig. 2, the full trainable unitary circuit can be modeled as a composition of \( N \) variational layers and can be denoted as \cite{garg2020advances}
\begin{equation}
    U{(\theta)} = U^{(N)}(\theta^{(N)})U^{(N-1)}(\theta^{(N-1)}) \cdots U^{(1)}(\theta^{(1)})
\end{equation}
where $U{(\theta)}$ denotes the unitary transformations of \ac{QNN}, $\theta=\{\theta^{(1)}, \theta^{(2)}, \cdots,  \theta^{(N)}\}$ represents the set of all trainable parameters, $U^{(\ell)}(\theta^{(\ell)})$, $\ell\in  \{1, 2, \dots, N\} $ is a parameterized quantum gate block with learnable parameters $\theta^{(\ell)}$ in $\ell$ layer, $N$ is the number of variational layers. After unitary transformations on the encoded state, we can get an output joint quantum state $\ket{\psi_{\mathbf{z}_{out}}}$ as follows \cite{ovalle2023quantum}
\begin{equation}
    \ket{\psi_{\mathbf{z}_{out}}}= U(\theta) \ket{\psi_{\mathbf{z}_{in}}}
\end{equation}
Then qubits are measured using the observable Pauli-Z, which yields a real-valued output vector $\mathbf{z}_{\text{out}}\in \mathbb{R}^{n_q}$ of QNN, which can be represented as \cite{ovalle2023quantum}
\begin{equation}
    \mathbf{z}_{\text{out}} = 
    \bra{ \psi_{\mathbf{z}_{out}}} Z^{\bigotimes{n_q}}\ket{\psi_{\mathbf{z}_{out}}}
\end{equation}
where $Z=\begin{bmatrix}
1 & 0 \\
0 & -1
\end{bmatrix}$ is the observable Pauli-Z. Finally, the $\mathbf{z}_{\text{out}}$ can be represented as 
\begin{equation}
    \mathbf{z}_{\text{out}} = [\langle Z_1 \rangle, \langle Z_2 \rangle, \dots, \langle Z_{n_q} \rangle]^T
\end{equation}
where each element \( \langle Z_i \rangle \in [-1, 1] \) represents the expected measurement outcome of a qubit. Based on $\mathbf{z}_{\text{out}}$, we feed it into a post linear layer that maps it to a \( K \)-dimensional logits vector \( \mathcal{S} \in \mathbb{R}^{K} \), 
 matching the number of users in this massive MIMO system. The linear layer can be represented as 
     \vspace{-0.08in}
\begin{equation}
    \mathcal{S} = \mathbf{W} \mathbf{z}_{\text{out}} + \mathbf{b}
\end{equation}
where \( \mathbf{W} \in \mathbb{R}^{{K} \times n_q} \) is a learnable weight matrix, \( \mathbf{b} \in \mathbb{R}^{K} \) is a learnable bias vector. The entries of \( \mathcal{S} \) represent the predicted scheduling score for each user, and larger values correspond to higher priority for downlink transmission. We apply a top-\( L \) selection policy based on a $logits$ vector \( \mathcal{S} \). The resulting binary user scheduling indicator vector $\boldsymbol{\xi}=[\mathbf{\xi}^1,\mathbf{\xi}^2,\dots,\mathbf{\xi}^K]$ is defined as
\begin{equation}
    \xi^l =
   \begin{cases}
1,  \text{if } l \in \text{Top}_L(\mathcal{S}) \\
0, \text{otherwise}
\end{cases}
\quad \text{for } l = 1, 2, \dots, K
\end{equation}
where \( \text{Top}_L(\mathcal{S}) \) returns the indices of the \( L \) largest elements in \( \mathcal{S} \). For the hybrid \ac{QNN} training, the final scheduling indicator vector is used to calculate the ergodic sum rate, with fairness as the reward function and the corresponding loss function.
\subsection{\ac{RL}-based Training Algorithm}
Algorithm 1 depicts the RL-based training algorithm that optimizes the hybrid \ac{QNN}'s performance efficiently and robustly. This model training algorithm takes the model $M$ that to be trained, two data loaders $\mathcal{D}_{train}$ and $\mathcal{D}_{val}$, and a manually defined epoch number $E$ as inputs, which will yield the outputs including a training reward value $R_{train}$ and two validation reward values $R_{val}^{det}, R_{val}^{sto}$. The algorithm utilizes an advanced policy gradient technique, enhanced by several key modifications: 
Initially, an \emph{$\epsilon$-greedy exploration strategy} is employed, balancing the discovery of novel, potentially superior scheduling policies (exploration) with the refinement of current productive ones (exploitation). The $\epsilon$ value decreases linearly over training epochs, progressively prioritizing precise exploitation over broad exploration as the model acquires more knowledge.
Additionally, a \emph{dynamic baseline $b$ guided by a cosine-annealed momentum $\alpha$} is included. This method significantly stabilizes the gradient updates by adapting its reference point to track recent and historical reward patterns. Such adaptability minimizes gradient variance, preventing erratic updates, and ensuring consistent and effective learning compared to a static baseline.

Furthermore, the algorithm features a \emph{direct matrix-based reward computation}. This approach optimizes the essential wireless sum-rate metric directly, utilizing the $\mathbf{G_{in}}$ interaction matrix to provide a relevant and immediate performance signal, aligning with the main objective to fine-tune network parameters for maximum impact.
To maintain training stability, \emph{gradient clipping and adaptive optimization} techniques are employed. These mechanisms prevent gradients from becoming excessively large or erratic, which could disrupt the learning process, thus ensuring a smoother and more reliable convergence.
Moreover, a \emph{dual validation strategy} is implemented. This covers both deterministic (optimal) and stochastic (practical) policy executions, offering a comprehensive evaluation of the model's performance across both idealized and real-world scenarios.
Finally, an optional \emph{checkpoint method} is included to enhance training robustness, allowing for the saving and restoration of model states along with their associated performance metrics.
Together, this precisely engineered RL algorithm addresses the inherent challenges of training quantum-classical hybrid models, especially their exposure to high-variance reinforcement learning settings, by fostering stability, efficiency, and exceptional performance.

\begin{algorithm}[!]
\small
\caption{\small Hybrid QNN Policy Training with RL}
\begin{algorithmic}[1]
\REQUIRE Model $M$, train loader $\mathcal{D}_{train}$, val loader $\mathcal{D}_{val}$, epochs $E$
\ENSURE Training rewards $R_{train}$, validation rewards $R_{val}^{det}, R_{val}^{sto}$
\STATE Initialize baseline $b \gets 0$, $\epsilon \gets 0.6$, $\alpha_{max} \gets 0.7$, $\alpha_{min} \gets 0.3$
\FOR{epoch $=1$ to $E$}
    \STATE Decay $\epsilon \gets \max(0.05, \epsilon - 0.55/E)$
    \STATE Compute $\alpha \gets \alpha_{min} + 0.5(\alpha_{max}-\alpha_{min})(1 + \cos(\pi \cdot \text{epoch}/E))$
    \STATE $M.\text{train}()$; $L \gets 0$; $R \gets 0$
    \FOR{$(x, \mathbf{G_in}) \in \mathcal{D}_{train}$}
        \STATE $logits \gets M(x)$ \COMMENT{Forward pass}
        \STATE $\pi \gets \sigma(logits)$ \COMMENT{Sigmoid probabilities}
        \STATE Sample policy $p \sim \text{Bernoulli}(\pi)$ with $\epsilon$-greedy exploration
        \STATE $r \gets \text{sumrate\_reward}(\mathbf{G_in}, p)$ \COMMENT{Matrix-based reward}
        \STATE $b \gets \alpha b + (1-\alpha)\text{mean}(r)$ \COMMENT{Update baseline}
        \STATE $\mathcal{L} \gets -\mathbb{E}[\log \pi \cdot (r - b)]$ \COMMENT{Policy gradient loss}
        \STATE $\nabla \mathcal{L}.\text{backward}()$; 
        \STATE $\text{optimizer.step}()$; $L \gets L + \mathcal{L}$; $R \gets R + r$
    \ENDFOR
    \STATE Evaluate $R_{val}^{det}, R_{val}^{sto}$ on $\mathcal{D}_{val}$ via top-$k$/sampling
    \STATE Save checkpoint with metrics (Optional)
\ENDFOR
\end{algorithmic}
\end{algorithm}
\section{Results}

We evaluate the proposed hybrid-\ac{QNN} algorithm's performance in various scenarios and compare it with a \ac{CNN} benchmark. Fig. 3 illustrates the training progress of the proposed hybrid-\ac{QNN} model in an RL setting, focusing on the combined loss metrics across different configurations. The average policy loss is reported for three distinct setups as shown in the legend. The 4-user-16-antenna and 8-user-32-antenna configurations performed stable and negative loss values, suggesting effective policy optimization. In contrast, the 12-user-64-antenna setup shows a significantly higher positive loss. Despite this, after 20 training iterations, the three lines converged around zero, with some ripples, demonstrating the robustness of our proposed model in handling different user-antenna configurations.

Fig. 4 reveals that, as the number of antennas increases, the sum rate generally improves. However, systems with fewer users achieve higher average sum-rates compared
to those with more users, since the interference management becomes more challenging with increasing user density, especially where antenna numbers are low. Despite this, when denser user configurations are used, our model
could achieve better enhancements while the antenna numbers
are scaled up. Thus, it can be safely stated that the denser the
user, the more sensitive our model is toward the scaled-up
antennas in terms of a higher increasing
rate of the sum-rate.

Fig. 5 examines the sum-rate performance across varying SNR levels from 0 to 25 dB for three different user-antenna configurations. As the \ac{SNR} increases, the sum rate improves for all configurations. As the number of antennas increases, the sum-rate generally improves for all three situations as well. Additionally, the overall sum-rate performance is proportional to the user-antenna configurations, resulting in the three lines increasing in parallel. These results demonstrate that better sum-rate performance is achieved not only with higher \ac{SNR} values but also facilitated by larger volumes of users and increased antenna numbers.

Fig. 6 compares the sum-rate performance of our proposed hybrid-\ac{QNN} model with a similarly architected \ac{CNN} benchmark under varying \ac{SNR} conditions for two system configurations. The proposed systems consistently outperform their CNN counterparts, particularly at higher \ac{SNR} levels. Both architectures exhibit improved sum-rates with increasing \ac{SNR}; however, the performance gap widens as the number of users and antennas increases, highlighting the superior scalability of \ac{QNN} in interference-rich environments. Therefore, \acp{QNN} could offer a promising alternative to classical approaches in next-generation wireless systems.
\begin{figure}[!]
    \centering
    \includegraphics[width=0.43\textwidth]{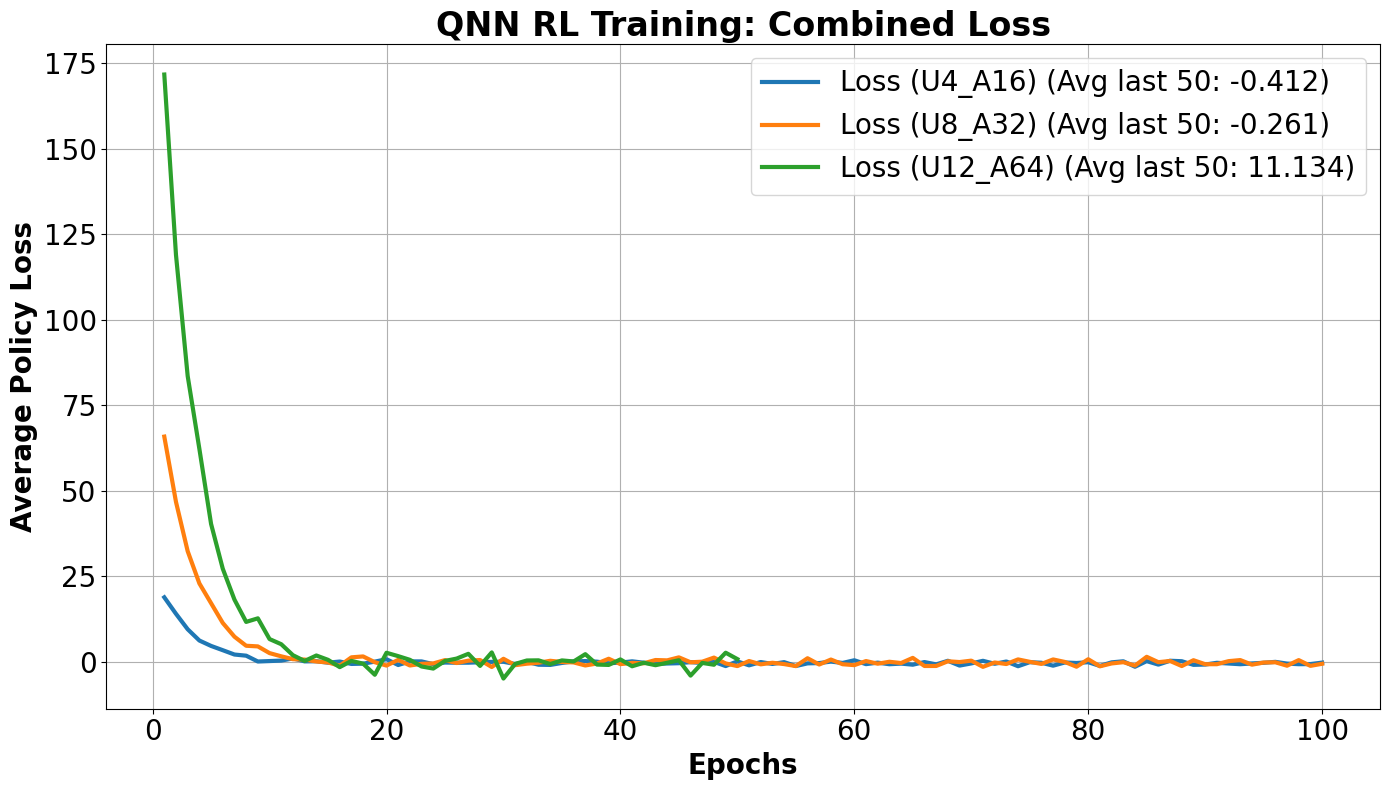}
    \caption{Training convergence of the hybrid-QNN scheduler showing policy loss minimisation, with multiple scenarios.}
    \label{fig:QNN_loss}
    \vspace{-0.1in}
\end{figure}
\begin{figure}[!]
    \centering
    \includegraphics[width=0.4\textwidth]{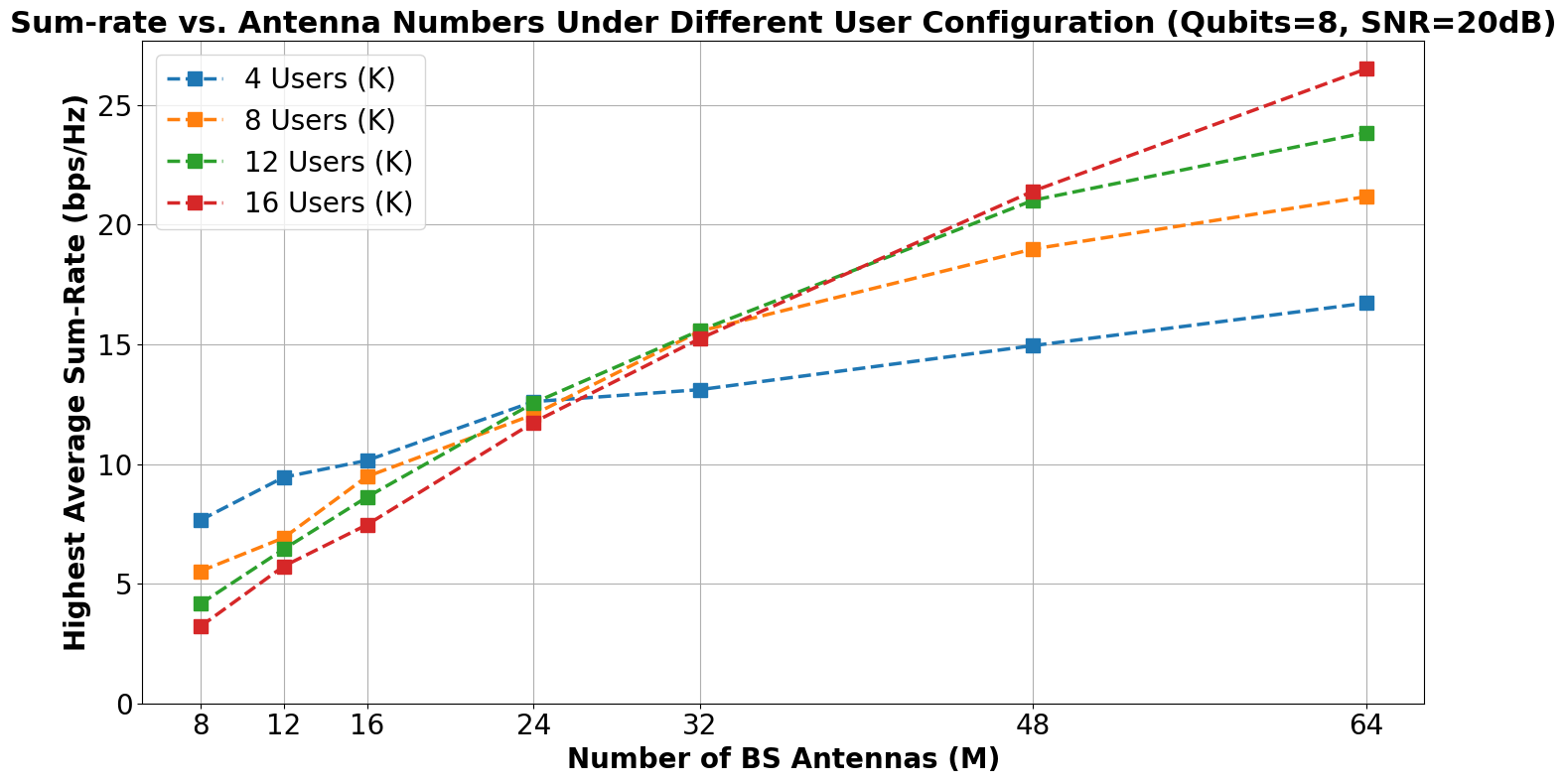}
    \caption{Sum-rate versus the number of antennas for different user configurations where qubits=8 and SNR=20dB.}
    \label{fig:sumrate_antennas}
    \vspace{-0.1in}
\end{figure}
\begin{figure}[!]
    \centering
    \includegraphics[width=0.4\textwidth]{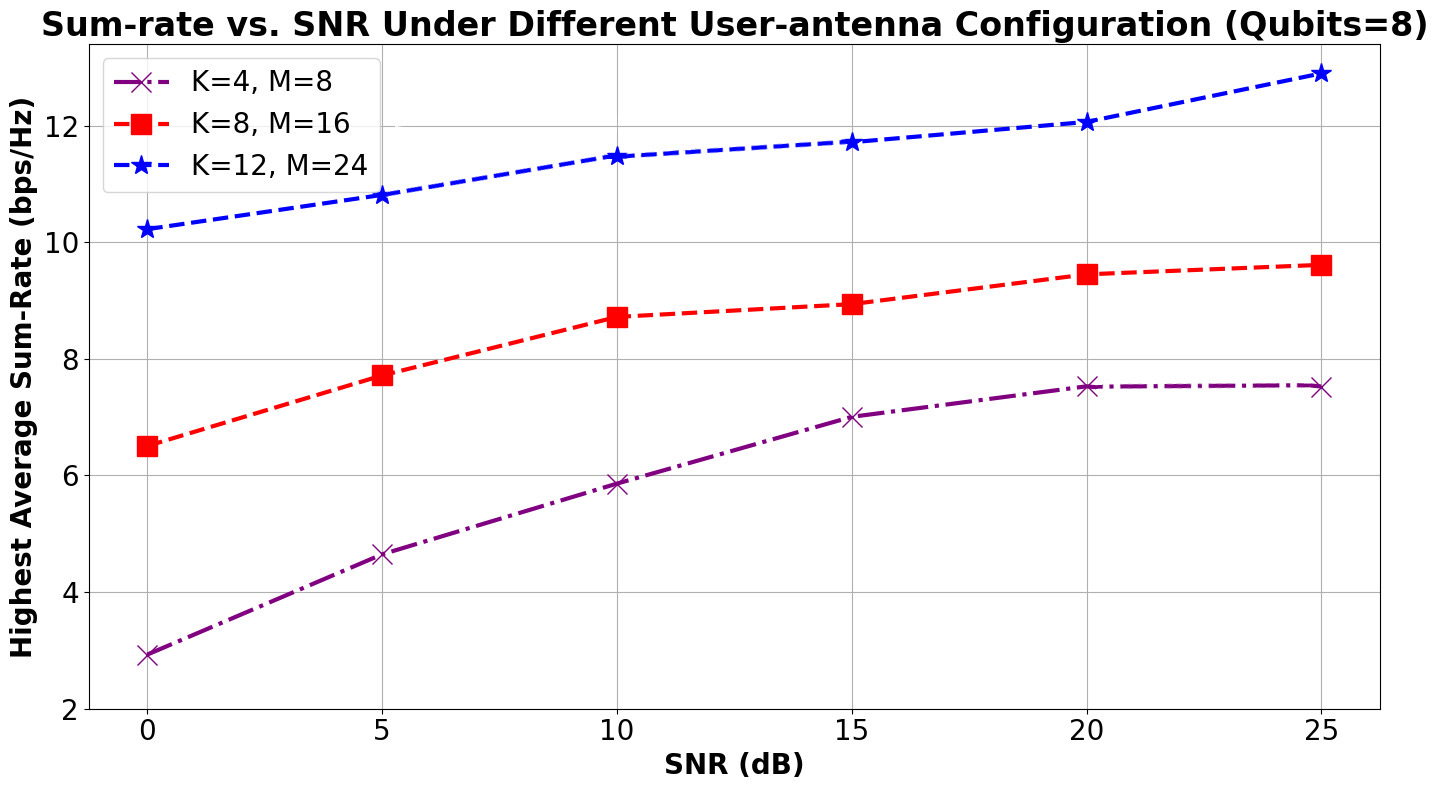}
    \caption{Sum-rate against SNR for different user-antenna configurations, where qubits=8.}
    \label{fig:sumrate_SNR}
    \vspace{-0.1in}
\end{figure}
\begin{figure}[!]
    \centering
    \includegraphics[width=0.4\textwidth]{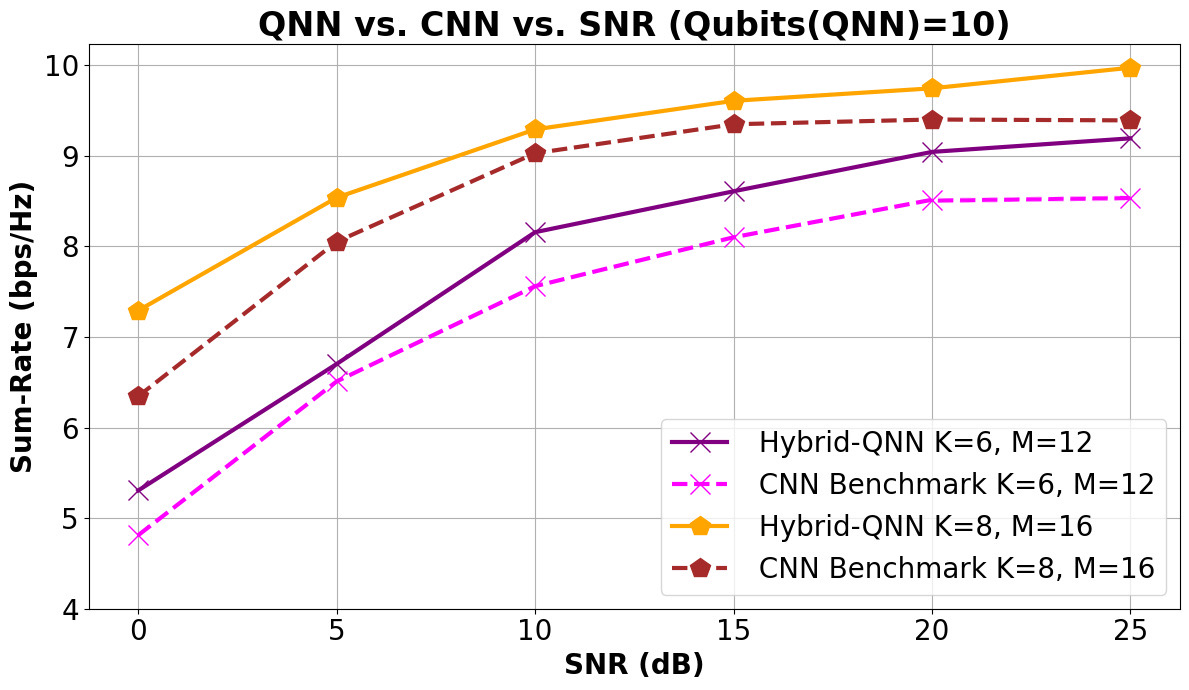}
    \caption{Sum-rate comparison between QNN and CNN across different SNR values where qubits=10 for QNN.}
    \label{fig:QNN_CNN_SNR}
    \vspace{-0.1in}
\end{figure}

\section{Conclusion}
This study investigates and proposes a hybrid-\ac{QNN} model for user scheduling in \ac{MU-mMIMO} systems, integrating conventional layers with a variational quantum circuit. The proposed model demonstrates rapid policy convergence and maximises ergodic sum-rate across various configurations, outperforming classical \ac{CNN} benchmarks, particularly in moderate-to-high \ac{SNR} regimes. It has also demonstrated robustness in the presence of high channel interference, which is often caused by dense user configurations. These findings establish hybrid-\acp{QNN} as a compelling approach for high-dimensional user scheduling and potentially, for resource allocation, in future communication systems.
\begin{acronym}
\acro{B5G}{Beyond 5G}
\acro{TDMA}{Time Division Multiple Access}
\acro{FDMA}{Frequency Division Multiple Access}
\acro{MIMO}{Multiple Input Multiple Output}
\acro{mMIMO}{massive Multiple Input Multiple Output} 
\acro{MU-mMIMO}{Multi-user Massive Multiple Input Multiple Output}
\acro{CSI}{Channel State Information}
\acro{CSIT}{Transmitter Channel State Information}
\acro{BS}{Base Station}
\acro{UT}{User Terminal}
\acro{AI}{Artificial Intelligence}
\acro{ML}{Machine Learning}
\acro{QNN}{Quantum Neural Networks}
\acro{QDL}{Quantum Deep Learning}
\acro{LoS}{Line of Sight}
\acro{NLoS}{Non Line-of-sight}
\acro{QoS}{Quality of Service}
\acro{WMMSE}{Weighted Minimum Mean Square Error}
\acro{DNN}{Deep Neural Networks} 
\acro{RL}{Reinforcement Learning}
\acro{DQN}{Deep Q-networks} 
\acro{PRO}{Policy Optimisation}
\acro{AWGN}{Additive White Gaussian Noise}
\acro{AoD}{Angle-of-departure}
\acro{DFT}{Discrete Fourier Transform}
\acro{ULA}{Uniform Linear Antenna Array}
\acro{NN}{Neural Networks} 
\acro{AP}{Access Point}
\acro{UE}{User Terminal}
\acro{SNR}{Signal-to-noise Ratio}
\acro{SINR}{Signal-to-interference-plus-noise Ratio}
\acro{SLNR}{Signal-to-leakage-plus-noise Ratio}
\acro{GPU}{Graphic Processing Unit}
\acro{PL}{Path Loss}
\acro{ANN}{Artificial Neural Networks} 
\acro{CNN}{Convolutional Neural Networks} 
\acro{GNN}{Graph Neural Networks} 
\acro{dB}{Decibel}
\acro{PF}{Proportional Fairness}
\acro{CNOT}{Controlled-NOT}
\end{acronym}
\bibliographystyle{IEEEtran}
\bibliography{reference}

\end{document}